\documentclass[a4paper,12pt]{article}
\pdfoutput=1 % if your are submitting a pdflatex (i.e. if you have
\usepackage{jheppub} % for details on the use of the package, please
\usepackage[T1]{fontenc} % if needed

\usepackage{amsmath}
\usepackage{graphicx,slashed,xcolor,multirow,bbold,mathtools,sidecap,tikz,bm,enumitem,booktabs,array}
\usepackage{verbatim}
\usepackage[normalem]{ulem}

\usepackage{tikz}
\usetikzlibrary{positioning,arrows}
\usetikzlibrary{decorations.pathmorphing}
\usetikzlibrary{decorations.markings}

\usepackage{subcaption}

\usepackage[compat=1.1.0]{tikz-feynman}
\newcommand{\beq}{\begin{equation}}
\newcommand{\eeq}{\end{equation}}
\newcommand{\bea}{\begin{eqnarray}}
\newcommand{\eea}{\end{eqnarray}}
\newcommand{\barr}{\begin{array}}
\newcommand{\earr}{\end{array}}

\def \st1{\widetilde t_1}
\def \mst1{m_{\st1}}
\def \sbot1{\widetilde b_1}

\def\bea{\begin{eqnarray}}
	\def\eea{\end{eqnarray}}

\usepackage{color}
\usepackage{colortbl}

\long\def\/*#1*/{}
\usepackage{color}
 \usepackage[normalem]{ulem}
\definecolor{darkgreen}{cmyk}{1,0,1,0.4}
\definecolor{darkred}{cmyk}{0,1,1,0.4}

\definecolor{lime}{HTML}{A6CE39}
\DeclareRobustCommand{\orcidicon}{\hspace{-2.1mm}
\begin{tikzpicture}
\draw[lime, fill=lime] (0,0) circle [radius=0.13] node[white] {{\fontfamily{qag}\selectfont \tiny \,ID}}; \draw[white, fill=white] (-0.0525,0.095) circle [radius=0.007]; 
\end{tikzpicture} \hspace{-3.2mm} }
\foreach \x in {A, ..., Z}{\expandafter\xdef\csname orcid\x\endcsname{\noexpand\href{https://orcid.org/\csname orcidauthor\x\endcsname} {\noexpand\orcidicon}}}

\title{Advancing Higgsino Searches by Integrating ML for Boosted Object Tagging and Event Selection}

\author[a,b]{Rameswar Sahu,\orcidA{}}
\author[a,b]{ Debabrata Sahoo,}
\author[a,b]{ Kirtiman Ghosh\orcidG{}}

\affiliation[a]{\footnotesize Institute of Physics, Bhubaneswar, Sachivalaya Marg, Sainik School Post, Bhubaneswar 751005, India}
\affiliation[b]{\footnotesize Homi Bhabha National Institute, Training School Complex, Anushakti Nagar, Mumbai 400094, India}

\emailAdd{rameswar.s@iopb.res.in}
\emailAdd{kirti.gh@gmail.com}
\emailAdd{debabrata.s@iopb.res.in}

\abstract{Higgsinos near the TeV mass range are highly motivated as they offer an elegant solution to the naturalness problem in the Standard Model. Extensive searches for such higgsinos within the framework of General Gauge Mediation (GGM) have been conducted by both the ATLAS and CMS collaborations. However, the sensitivity of these searches in the hadronic channel remains limited, primarily due to the reliance on traditional substructure-based techniques for fat jet identification. In this work, we present a novel search strategy that leverages graph neural networks (GNNs) to improve the characterization of fat jets originating from $W/Z/h$ bosons, top quarks, and QCD-initiated light quarks and gluons. The GNN scores, combined with a boosted decision tree (BDT) classifier, enhance signal-background discrimination, offering a significant improvement in sensitivity for higgsino searches at the LHC.   }


\begin{document} 

\tikzset{
  vector/.style={decorate, decoration={snake,amplitude=.4mm,segment length=2mm,post length=1mm}, draw},
  tes/.style={draw=black,postaction={decorate},
    decoration={snake,markings,mark=at position .55 with {\arrow[draw=black]{>}}}},
  provector/.style={decorate, decoration={snake,amplitude=2.5pt}, draw},
  antivector/.style={decorate, decoration={snake,amplitude=-2.5pt}, draw},
  fermion/.style={draw=black, postaction={decorate},decoration={markings,mark=at position .55 with {\arrow[draw=blue]{>}}}},
  fermionbar/.style={draw=black, postaction={decorate},
    decoration={markings,mark=at position .55 with {\arrow[draw=black]{<}}}},
  fermionnoarrow/.style={draw=black},
  % gluon/.style={decorate, draw=black,decoration={coil,amplitude=4pt, segment length=5pt}},
  scalar/.style={dashed,draw=black, postaction={decorate},decoration={markings,mark=at position .55 with {\arrow[draw=blue]{>}}}},
  scalarbar/.style={dashed,draw=black, postaction={decorate},decoration={marking,mark=at position .55 with {\arrow[draw=black]{<}}}},
  scalarnoarrow/.style={dashed,draw=black},
  electron/.style={draw=black, postaction={decorate},decoration={markings,mark=at position .55 with {\arrow[draw=black]{>}}}},
  bigvector/.style={decorate, decoration={snake,amplitude=4pt}, draw},
  particle/.style={thick,draw=blue, postaction={decorate},
    decoration={markings,mark=at position .5 with {\arrow[blue]{triangle 45}}}},
  % fermion/.style={thick,draw=blue, postaction={decorate},
  %   decoration={markings,mark=at position .56 with {\arrow[blue]{triangle 45}}}},
  gluon/.style={decorate, draw=black,
    decoration={coil,aspect=0.3,segment length=3pt,amplitude=3pt}}
}

\maketitle
\flushbottom

\section{Introduction}
\label{sec:intro}
As a solution to the \textit{naturalness} problem, Supersymmetry (SUSY) has been advocated and studied extensively in the literature \cite{Sakai:1981gr,Dimopoulos:1981yj,Ibanez:1981yh,Dimopoulos:1981zb}. In "natural" supersymmetric theories, additional diagrams from the superpartners of the SM fermions and gauge boson help in canceling the divergent part of the radiative correction to the Standard Model (SM) Higgs boson mass. Over the past few decades, such SUSY models have managed to retain the attention of both phenomenologists and experimentalists because of their attractive predictions that can be tested in the current and upcoming collider experiments. \\

Notably, these theories predict that the top, bottom, gluon, and Higgs superpartners must be relatively light \cite{Barbieri:1987fn}, enabling their production at the energies available at the Large Hadron Collider (LHC). Among these, stops, sbottoms, and gluinos—owing to their color charge—can be produced abundantly at the LHC. Consequently, the ATLAS and CMS experiments have placed stringent constraints on their masses \cite{ATLAS:2024lda,CMS:2024rvf,Adam:2021rrw}. However, the study of pure higgsinos presents unique challenges \cite{SUSY-2017-02, CMS-SUS-20-004,SUSY-2018-02, SUSY-2018-41, SUSY-2018-05, ATLAS:2024zzz}. In SUSY, pure higgsinos form a triplet comprising two neutralinos ($\tilde{\chi}_1^0,\tilde{\chi}_2^0$) and a chargino ($\tilde{\chi}_1^{\pm}$) with nearly degenerate masses. At the LHC, higgsino pairs can be produced through four primary channels with experimentally accessible cross-sections: $\tilde{\chi}_1^0 \tilde{\chi}_2^0$, $\tilde{\chi}_1^0\tilde{\chi}_1^{\pm}$, $\tilde{\chi}_2^0\tilde{\chi}_1^{\pm}$, and $\tilde{\chi}_1^{\pm}\tilde{\chi}_1^{\pm}$. In the Minimal Supersymmetric Standard Model (MSSM) with conserved R-parity \cite{Farrar:1978xj}, the lightest neutralino is stable and manifests as missing transverse momentum in the detector. Higgsino pair production at the LHC is therefore expected to result in soft quarks and leptons from the decays of $\tilde{\chi}_2^0$ and $\tilde{\chi}_1^{\pm}$ to $\tilde{\chi}_1^0$, along with some amount of missing transverse momentum. Being very soft, these quarks and leptons are very difficult to detect, and therefore, the experiments usually rely on hard partons from initial state radiations (ISR) to probe these higgsinos. At LHC, the ATLAS experiment has been able to exclude $\tilde{\chi}_2^0$ mass below 193 GeV at 95\% C.L. \cite{ATLAS:2019lng}, for a 9.3 GeV mass splitting between the two neutralinos. Similarly, by looking into final states containing two or three low-momentum leptons (with at least one opposite sign pair) and large missing transverse energy induced by the jets originating from the ISR, the CMS collaboration has excluded higgsino mass up to 205 GeV at 95\% C.L. \cite{CMS:2021edw} for a 7.5 GeV mass splitting between the two higgsino like neutralinos. The primary reason for such low sensitivities is the soft leptons and jets originating from the decay of the higgsino states. However, there are extensions of the MSSM where this problem can be avoided, and such theories can provide a framework for exploring pure higgsinos. In this work, we plan to study one such scenario.\\

Though a compelling theoretical framework that provides solutions to many existing problems of the SM \cite{Golfand:1971iw, Volkov:1973ix, Wess:1974tw, Dimopoulos:1981wb, Dimopoulos:1981zb, Dimopoulos:1981yj, Sakai:1981gr, Ibanez:1981yh, Einhorn:1981sx, Ellis:1983ew, Goldberg:1983nd, Mizuta:1992qp, Cirelli:2007xd, Feng:2000gh, Hisano:2006nn}, supersymmetry cannot be an exact symmetry of nature and must be broken at least softly at some high energy scale \cite{Dimopoulos:1981zb, Girardello:1981wz, Sakai:1981gr}. Over the past several decades, physicists have extensively studied several mechanisms \cite{Nilles:1982ik, Chamseddine:1982jx, Nath:1983aw, Barbieri:1982eh, Cremmer:1982vy, Ibanez:1982ee, Nilles:1982dy, Randall:1998uk, Giudice:1998xp, Dine:1993yw, Dine:1994vc, Dine:1995ag} of SUSY breaking. Gauge-mediated supersymmetry breaking (GMSB) \cite{Dine:1993yw, Dine:1994vc, Dine:1995ag} or its more general form, general gauge mediation (GGM) \cite{Meade:2008wd, Buican:2008ws} is one such framework. The GGM framework assumes the breaking of SUSY in a hidden sector, which is then communicated to the visible sector through some messenger fields that only have SM gauge interaction. In the limit of vanishing gauge couplings, these two sectors completely decouple. Additionally, since the gauge interactions are flavor blind, there is no SUSY flavor problem. If we promote SUSY to be a local symmetry, the particle associated with the spontaneous breaking of SUSY gets absorbed by the fermionic superpartner of the graviton (the gravitino $\tilde{G}$). As the scale of SUSY breaking is much smaller than that of the Plank scale, the gravitino becomes nearly massless (see Ref. \cite{Ajaib:2012vc} for details) and the lightest supersymmetric particle (LSP) of the theory. In such theories, the collider signature of pure higgsinos becomes interesting, as they can now decay into the gravitino along with other SM particles, even when the R-parity is conserved.\\

The search for pure higgsinos in theories of general gauge mediation has been done extensively by both the ATLAS \cite{ATLAS:2021yyr, ATLAS:2021yqv, ATLAS:2024tqe, ATLAS:2018rns, ATLAS:2020qlk, ATLAS:2018tti} and CMS \cite{CMS:2021cox, CMS:2024gyw, CMS:2017moi, CMS:2020bfa, CMS:2017bki, CMS:2017nin, CMS:2022vpy} collaboration. Most of these studies assume the mass difference among the higgsino triplet to be very small so that the SM particles resulting from the decay of $\tilde{\chi}_2^0/\tilde{\chi}_1^{\pm}$ to $\tilde{\chi}_1^0$ are very soft and hence non-detectable at the LHC. Additionally, they also assume the coupling between $\tilde{\chi}_2^0/\tilde{\chi}_1^{\pm}$ and the gravitino to be very small so that $\tilde{\chi}_2^0/\tilde{\chi}_1^{\pm}$ always decay to $\tilde{\chi}_1^0$. At the same time, the coupling between $\tilde{\chi}_1^0$ and the gravitino is large enough to allow prompt decay of the lightest neutralino. Since the lightest neutralino can have two possible decay modes, i.e., $\tilde{\chi}_1^0\rightarrow h \tilde{G}$ (where $h$ is the SM-like Higgs boson) and $\tilde{\chi}_1^0 \rightarrow Z \tilde{G}$, we can expect several possible final state topologies based on the leptonic and hadronic decays of the $Z$ and $h$ boson. \\

In Ref. \cite{ATLAS:2021yyr}, the ATLAS collaboration investigated final states containing four or more charged leptons, excluding higgsino masses in the GGM framework up to 540 GeV at a 95\% confidence level. Similarly, the CMS collaboration analyzed events with three or four leptons \cite{CMS:2021cox}, excluding higgsino masses up to 600 GeV (400 GeV) assuming a 100\% (50\%) decay branching ratio of higgsinos through the $Z$ channel, and up to 200 GeV for a 100\% decay branching ratio through the $h$ channel. ATLAS also explored final states featuring two boosted hadronically decaying bosons and large missing transverse momentum \cite{ATLAS:2021yqv}, excluding higgsino masses in the range of 450–940 GeV (500–850 GeV) for BR($\tilde{\chi}_1^0\rightarrow Z \tilde{G}$) = 100\% (50\%). Furthermore, in Ref. \cite{ATLAS:2024tqe}, ATLAS probed final states with at least three b-jets and significant missing transverse momentum, excluding higgsino masses between 130 GeV and 940 GeV under the assumption of a 100\% decay of the lightest neutralino into a Higgs boson and gravitino. The CMS collaboration extended these searches by incorporating multiple final state topologies, including both leptonic and hadronic channels \cite{CMS:2024gyw}. This comprehensive approach excluded higgsino masses up to 1025 GeV when the lightest neutralino exclusively decays into a Higgs boson and gravitino.\\

Clearly, the searches focusing on hadronic final state topologies have a better reach than the leptonic ones due to the higher hadronic decay branching ratio of the SM $Z/h$ bosons, which significantly enhances the overall signal yield. However, these hadronic search strategies face two notable challenges. First, analyses targeting the Higgs decay mode of the higgsino rely on multiple $b$-tagged jets, which inherently reduce the signal yield. Second, searches employing boosted objects predominantly use traditional jet substructure techniques to identify jets originating from final-state $Z/h$ bosons. Recent advancements in machine learning (ML) have consistently demonstrated that modern, ML-based jet taggers often outperform these conventional approaches. To address this limitation, we propose a detailed investigation employing a graph neural network (GNN) based multi-class classifier for efficient identification of the parent particles of these fat jets. We train this multiclass classifier to characterize fat jets originating from the hadronic decays of SM $W/Z/h$ bosons, top quarks, and QCD-initiated light quarks and gluons in terms of four scores, namely the top, $W/Z$, Higgs, and QCD score. Unlike traditional machine learning (ML)-based approaches that assign fat jets to specific classes with fixed tagging efficiencies, our methodology leverages the GNN scores as high-level variables instead of directly tagging the jets. These scores, combined with other carefully constructed variables characterizing the signal events, are fed into a boosted decision tree (BDT) classifier for signal and background discrimination. Our strategy reduces the dependency on multiple $b$-tagging to reduce SM backgrounds. This integrated approach effectively reduces background contamination in the signal regions and has the potential to significantly enhance the overall sensitivity of the search.

%In this paper, we propose a novel search strategy that utilizes a graph neural network-based multiclass classifier to improve the identification of boosted, hadronically decaying $W/Z/h$ bosons, top quarks, and QCD-initiated light quarks and gluons. To ensure the completeness of our strategy, we have looked into both hadronic and leptonic final states. Unlike the above-mentioned experimental studies that take into account only the $bb$ decay mode of the Higgs boson, we have also included the $cc,~ WW, ~\rm{and}~ ZZ$ decay modes. The better identification of boosted fat jets and incorporation of all dominant decay modes of the Higgs boson into our analysis helps significantly improve the reach of our strategy. \\

The rest of the paper is as follows: In Section \ref{sec:model}, we briefly discuss the simplified GGM model under study. Section \ref{sec:collider_analysis} contains the details of our analysis, including the simulation of signal/background events, object reconstruction, and event selection. In Section \ref{sec:result}, we present our findings before concluding our work in Section \ref{sec:concl}.

\section{Model definition}
\label{sec:model}
The model considered in our analysis is a simplified version of the GGM-type SUSY breaking scenarios with conserved R-parity. The model features a nearly massless gravitino ($\tilde{G}$) as the lightest SUSY particle (LSP) and an almost mass degenerate higgsino triplet consisting of two neutralinos ($\tilde{\chi}_1^0, \tilde{\chi}_2^0$) and one chargino ($\tilde{\chi}_1^{\pm}$). As discussed in the introduction, our model assumes the coupling of $\tilde{\chi}_2^0$ and $\tilde{\chi}_1^{\pm}$ to the gravitino to be small so that their direct decay to $\tilde{G}$ is highly suppressed. At the same time, the coupling between $\tilde{\chi}_1^0$ and $\tilde{G}$ is assumed to be large enough to allow prompt decay of the lightest neutralino. All other SUSY particles are assumed to decouple at a high mass.\\ 

\begin{figure}
    \centering
    \includegraphics[width=0.4\linewidth]{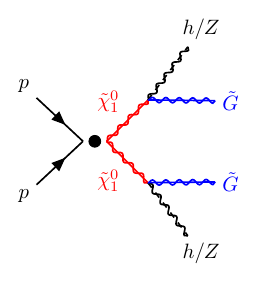}
    \caption{Feynman diagram depicting the production of higgsino pairs and their subsequent decay into $h/Z$-mediated final states.}
    \label{fig:feyn_diag}
\end{figure}

R-parity being conserved SUSY particles can only be produced in pairs, and their decay must contain an odd number of SUSY particles in the final state. Considering the particle spectrum of our model, the only phenomenological relevant production channels are: $p p \rightarrow \tilde{\chi}_1^0 \tilde{\chi}_2^0$, $p p \rightarrow \tilde{\chi}_1^0 \tilde{\chi}_1^{\pm}$, $p p \rightarrow \tilde{\chi}_2^0 \tilde{\chi}_1^{\pm}$, and $p p \rightarrow \tilde{\chi}_1^{\pm} \tilde{\chi}_1^{\pm}$. Following production, the $\tilde{\chi}_2^0$ and $\tilde{\chi}_1^{\pm}$ decay to the lightest neutralino along with SM quarks/leptons. However, due to the near mass degeneracy of the charginos and neutralinos, the SM quarks and leptons resulting from these decays are very soft and are of no phenomenological significance. The lightest neutralino ($\tilde{\chi}_1^0$) subsequently decays into the LSP gravitino, accompanied by an SM like Higgs or $Z$ boson \footnote{ Note that the $Z$ mediated decay channel of the lightest neutralino helps in accounting for any bino/wino component in the lightest neutralino.}. Consequently, the collider signature of our model will effectively look like the diagram depicted in figure \ref{fig:feyn_diag}. \\

Our simplified implementation of GGM has only two phenomenological important parameters: The higgsino mass $M_{\tilde{\chi}_1^0}$ and the branching ratio of the lightest neutralino decay to $h + \tilde{G}$ or $Z + \tilde{G}$ final states. We have varied these two parameters independently in our final analysis. The higgsino mass is scanned across a range of 800 to 1500 GeV, with a step size of 50 GeV. For simplicity, we consider three discrete values for the branching ratio of the lightest neutralino to the Higgs boson and gravitino: 100\%, 50\%, and 25\%. Other fixed parameters include the gravitino mass, set at 1 eV, the ratio of vacuum expectation values of the two Higgs doublets ($\rm{tan} \beta$), fixed at 1.5, and the masses of all remaining sparticles, fixed at 5 TeV.\\

\section{Collider Phenomenology}
\label{sec:collider_analysis}
This section will focus on our analysis methodology, which includes signal and background event generation, object reconstruction, and event selection. As outlined in the previous section, the simplified GGM model considered in this work favors the production of higgsinos via four production channels. However, considering the mass degeneracy between the charginos and neutralinos, all these channels will result in the same final states at the LHC. Therefore, for simplicity, we only considered the direct production of neutralino pairs (see Figure \ref{fig:feyn_diag}) at the LHC while generating signal events for our analysis. The dominant SM background processes include single and multi-top production with/without additional vector bosons and mono, di, tri, and tetra boson processes. \\

\subsection{Event generation and Object Reconstruction}
\label{sec:obj_recon}
To generate the signal events, we have modified the default MSSM model file of \texttt{SARAH-4.14.5} \cite{SARAH1, SARAH2} to incorporate the gravitino-mediated decays of the charginos and neutralinos. The SUSY particle spectrum has been generated with \texttt{SPheno-4.0.4} \cite{SPheno1, SPheno2}. Signal events up to two additional partons are generated at leading order \texttt{MadGraph5-aMC@NLO} \cite{MadGraph}. The signal cross-sections at the NLO+NLL accuracy are calculated using \texttt{Resummino-3.1.1} \cite{Resummino}. The SM backgrounds considered in our analysis includes $t\bar{t}$, $ttV$, $ttVV$, $VV$, $VVV$, $VVVV$ (V = $w^+/w^-/z$), $W+j$, $Z+j$, and $tw$. All background processes up to two additional partons are generated at leading order in \texttt{MadGraph5-aMC@NLO} \cite{MadGraph}. The LO background cross-sections are scaled with appropriate NLO K-factors \cite{Muselli:2015kba, Catani:2009sm, Balossini:2009sa, Broggio:2019ewu, Kidonakis:2015nna, Campbell:2011bn, Shen:2015cwj}. Decays of the unstable particles, showering, fragmentation, and hadronization are simulated in \texttt{PYTHIA 8.2} \cite{Pythia8.2}. To incorporate detector effects in our analysis, we have used the fast detector simulator \texttt{Delphes-3.4.2} \cite{Delphes} with the default ATLAS card.\\

Jets are reconstructed using the \textit{anti-$k_T$} algorithm \cite{Anti-KT} with \texttt{FastJet} \cite{FastJet}. Considering the mass of the neutralinos and the SM $Z/h$ bosons, we expect these bosons to be highly boosted. Therefore, the jets produced from the decay of these bosons will be highly collimated, and it may not be possible to reconstruct these jets as individual final-state objects. Instead, it is more economical to reconstruct these bosons as a single large-radius fat jet. For our analysis, we used both small radius jets (reconstructed with a radius parameter $R=0.4$) and fat jets (reconstructed with a radius parameter $R=1.2$). The $R=0.4$ jets are required to have a $p_T$ greater than 25 GeV and $|\eta|$ less than 2.8. To identify jets originating from bottom quarks, we used Delphes' default b-tagging algorithm. The fat jets are required to have $p_T > 300$ GeV and $|\eta| <$ 2.5. Since we expect these jets to originate from the boosted $Z/h$ bosons, we have demanded their invariant mass to be higher than 70 GeV. This condition can be considered as a pre-selection requirement on the fat jets which can help in reducing contribution from QCD jets. Once these fat jets are reconstructed, we pass them through a GNN-based multiclass classifier to compute the characteristics of the fat jets in terms of the four GNN scores, namely the top, W/Z (V), h, and QCD score. A detailed discussion on this classifier is presented in Appendix \ref{sec:gnn_model}. This multiclass classifier is trained to characterize fat jets belonging to four classes: top, V (W/Z), h, or QCD jets. Note that fat jets originating from $W$ and $Z$ bosons are placed in a single class due to their close mass and almost identical final state decay topologies. Also, note that in our analysis, we are not removing the small radius jets that reside inside the fat jets. The definition of signal regions and construction of kinematic variables are done carefully so that the overlap between light and fat jets does not cause any problems.\\

Electron and muon candidates are required to have $p_T >$ 15 GeV and $\eta <$ 2.5. Electrons within the barrel and endcap (i.e., 1.37 $<|\eta|<$ 1.52) region of the electromagnetic calorimeter are removed from our analysis. Following Ref. \cite{ATLAS:2019jvq}, simple track and tower-based isolation criteria are imposed on the selected electron and muon candidates. Finally, a procedure based on the prescription of the ATLAS collaboration \cite{ATLAS:2016poa, ATLAS:2020uiq} is implemented to avoid possible double counting among selected leptons and small radius jets. The missing energy \footnote{Note that the fat jets are not included in the definition of missing energy, as the fat jet constituents are already present inside the light jets. This is because our analysis does not incorporate any criteria to remove this ambiguity.} of the system is defined as the magnitude of the negative vector sum of the transverse momentum of all visible final state objects (excluding the fat jets) and the tracks not associated with any such objects. 

\subsection{Event Selection}
\label{sec:even_select}
For the final analysis, we define seven carefully constructed signal regions (SRs) tailored to the expected composition of leptons ($N_l$), fat jets ($N_{fj}$), small radius light jets ($N_j$), and $b$-jets ($N_b$) in the signal events. Throughout our analysis, $N_j$ includes both the light flavor jets and $b$ jets. For clarity, we use the symbol $j$ to represent all small-radius jets, including $b$-jets, and the symbol $b$ specifically to denote small-radius $b$-tagged jets. The details of these SRs are summarized in Table \ref{tab:signal_regions}. Each signal region is mutually exclusive, ensuring that the majority of signal events are effectively captured within one of these regions.

To suppress the SM background contributions, we apply a set of robust pre-selection criteria. These include thresholds on the transverse momentum ($p_T$) of the two leading small-radius jets (including b-jets), their rapidity, the missing transverse energy ($\not\!p_T$), missing energy significance ($\not\!p_T/\sqrt{H_T}$), effective mass ($M_{eff}$), and the azimuthal separation between the two leading light jets and $\not\!p_T$. Here, $H_T$ is defined as the scalar sum of the transverse momenta of all visible final-state particles, while $M_{eff}$ is the sum of $H_T$ and $\not\!p_T$. 

It is crucial to emphasize that the first five signal regions in Table~\ref{tab:signal_regions} are specifically designed to target fully hadronic final states. As a result, the SM multijet background poses a significant challenge for these regions due to its exceptionally large cross-section. Simulating such a background is computationally challenging. However, previous ATLAS searches~\cite{ATLAS:2021yqv, ATLAS:2020syg} have established robust criteria that can effectively eliminate the multijet background. The preselection requirements outlined in Table~\ref{tab:signal_regions} are directly inspired by these ATLAS studies. Accordingly, we have not included the multijet background in our analysis, as it is expected to be negligible after applying these preselection cuts.

\begin{table}[htb!]
	\begin{center}
		\begin{tabular}{l|c|c|c|c|c|c|c}
			\toprule 
             Cuts& 0l1f0b& 0l1f1b & 0l1f2b & 0l1f3b & 0l2f & 1l1f & 1l2f\\
            \midrule
             $N_{l}$  & 0    & 0 & 0&  0& 0 & $\geq$ 1 & $\geq$ 1\\

             $N_{fj}$         & 1      & 1 & 1    & 1   & $\geq$ 2 & 1   & $\geq$ 2  \\

             $N_j$         & $\ge 2$  & $\ge 2$   & $\ge 2$&  $\ge 2$  & $\ge 2$   & $\ge 2$   & $\ge 2$ \\

             $N_b$      & 0  & 1   & 2 &  $\ge 3$  & -- & -- & -- \\
             \midrule
             $p_T (j1)$ (GeV)    & $>200$  & $>200$ & $>200$&  $>200$& $>200$ & $>100$ & $>100$ \\

              $p_T (j2)$ (GeV)   & $>100$    & $ >100 $ & $>100$&  $ >100 $&$>100$ &$>50$ &$>50$\\

              $\eta (j_{1,2})$   & $<2$    & $<2$ & $<2$ &  $<2$ & $<2$ &$<2$ &$<2$\\

             $\not\!p_T$ (GeV)   & $>300$     & $>300$ & $>300$ &  $>300$ & $>300$ & $>300$ & $>300$ \\

             $\Delta \phi(j_1,\not\!p_T)$   & $>0.8$     & $>0.8$ & $>0.8$ &  $>0.8$ & $>0.8$ & $>0.4$ & $>0.4$ \\

             $\Delta \phi(j_2,\not\!p_T)$   & $>0.8$     & $>0.8$ & $>0.8$ &  $>0.8$ & $>0.8$ & $>0.2$ & $>0.2$ \\

             $M_{eff}$ (GeV)   & $>1000$     & $>1000$ & $>1000$ &  $>1000$ & $>1000$ & $>800$ & $>800$ \\

             $\not\!p_T/\sqrt{H_T}$ ($\sqrt{\rm{GeV}}$)   & $>100$     & $>100$ & $>100$ &  $>100$ & $>100$ & $>10$ & $>10$ 	\\
			\bottomrule
		\end{tabular} 
		\caption{\label{tab:signal_regions} Pre-selection cuts for the seven signal regions implemented in our analysis.}
	\end{center}
\end{table} 

For each signal region, we have carefully identified kinematic variables that effectively capture the distinctive features of the signal events. The complete list of these variables is provided in Appendix \ref{sec:kinametic_variables}; however, we discuss some key variables here for clarity. As noted earlier, instead of assigning discrete class labels to the fat jets, our analysis utilizes the GNN-derived class scores to characterize the fat jets. This approach yields three essential variables: the $t$-score ($S_{t,i}$), $v$-score ($S_{v,i}$), and $h$-score ($S_{h,i}$), where the index $i$ runs over signal-region fat jets that satisfy the $p_T$ and invariant mass criteria. Since we do not expect top fat jets in the signal events and Higgs-initiated fat jets in the dominant background events, we expect the t-score and h-score to be highly effective in discriminating signal events from the background. Additionally, for the mass range of higgsinos considered in our study, we expect higher values of $H_T$ and $M_{eff}$ (see Section \ref{sec:even_select}) for signal events. 

In the signal region designed to capture fully hadronic final state topologies, we expect at least 3-4 high $p_T$ small radius jets from the decay of final state $Z/h$ bosons. Therefore, a variable like $(p_{T,j_1}+p_{T,j_2})/H_T$ is expected to take a smaller value for the signal events. Signal events are also characterized by higher missing transverse energy, making both $\not\!\!p_T$ and missing energy significance ($\not\!p_T / \sqrt{H_T}$) crucial for distinguishing signal from background contributions. In addition to these, we have employed several other kinematic variables that optimally capture the characteristics of the signal events (see Appendix \ref{sec:kinametic_variables}). These variables serve as inputs to boosted decision tree (BDT) classifiers, with one BDT trained for each signal region, enabling effective discrimination between signal and background events across the signal regions.

The training and evaluation of the classifiers are conducted using the Python package XGBoost~\cite{Chen_2016}. Details of the relevant BDT hyperparameters are discussed in Appendix~\ref{sec:hype_params}. As outlined in Section~\ref{sec:model}, our analysis considers three distinct branching ratio (BR) scenarios for the decay of the lightest neutralino into the $h + \tilde{G}$ final state. We treat these three signal scenarios independently while training the BDT classifiers. For each BR scenario, the neutralino mass is varied between 800 GeV and 1.5 TeV in steps of 50 GeV. One million signal events are generated for each neutralino mass, with half allocated for training and the remaining half reserved for testing the BDT classifiers. 

Before using the events for training/evaluation of the classifiers, we categorize them into one of the seven signal regions discussed above and pass them through the preselection requirements. Only events passing the preselection requirements are used to train and evaluate the classifiers. For each of the three signal scenarios, we have trained seven classifiers for the seven signal regions separately. While training the classifiers, signal events corresponding to different values of neutralino masses are combined to form the signal class. For the background class, events are generated from the background processes outlined in Section \ref{sec:collider_analysis}. For proper training of the classifier, we have generated enough events so that after preselection, about 5k background events from each background process remain for training the classifier in each signal region. These background events are then weighted by their respective cross-sections to reflect their realistic contributions in an experimental setting. For testing the classifiers, we have ensured that the number of background events generated is equal to or greater than the number of events expected in 200 $fb^{-1}$ integrated luminosity of LHC data.

\section{Results}
\label{sec:result}
\begin{figure}
    \centering
    \includegraphics[width=0.3\linewidth]{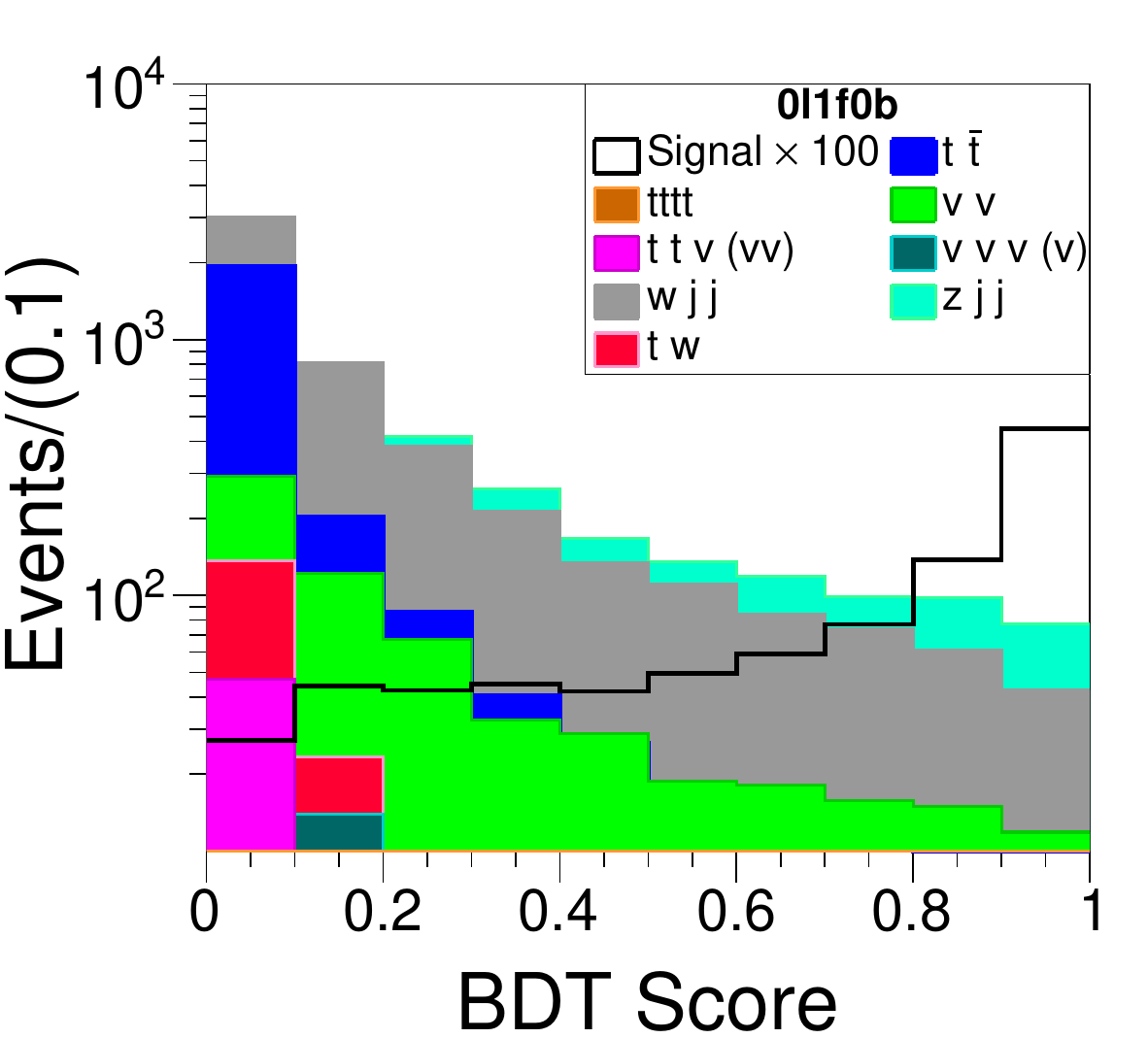}
    \includegraphics[width=0.3\linewidth]{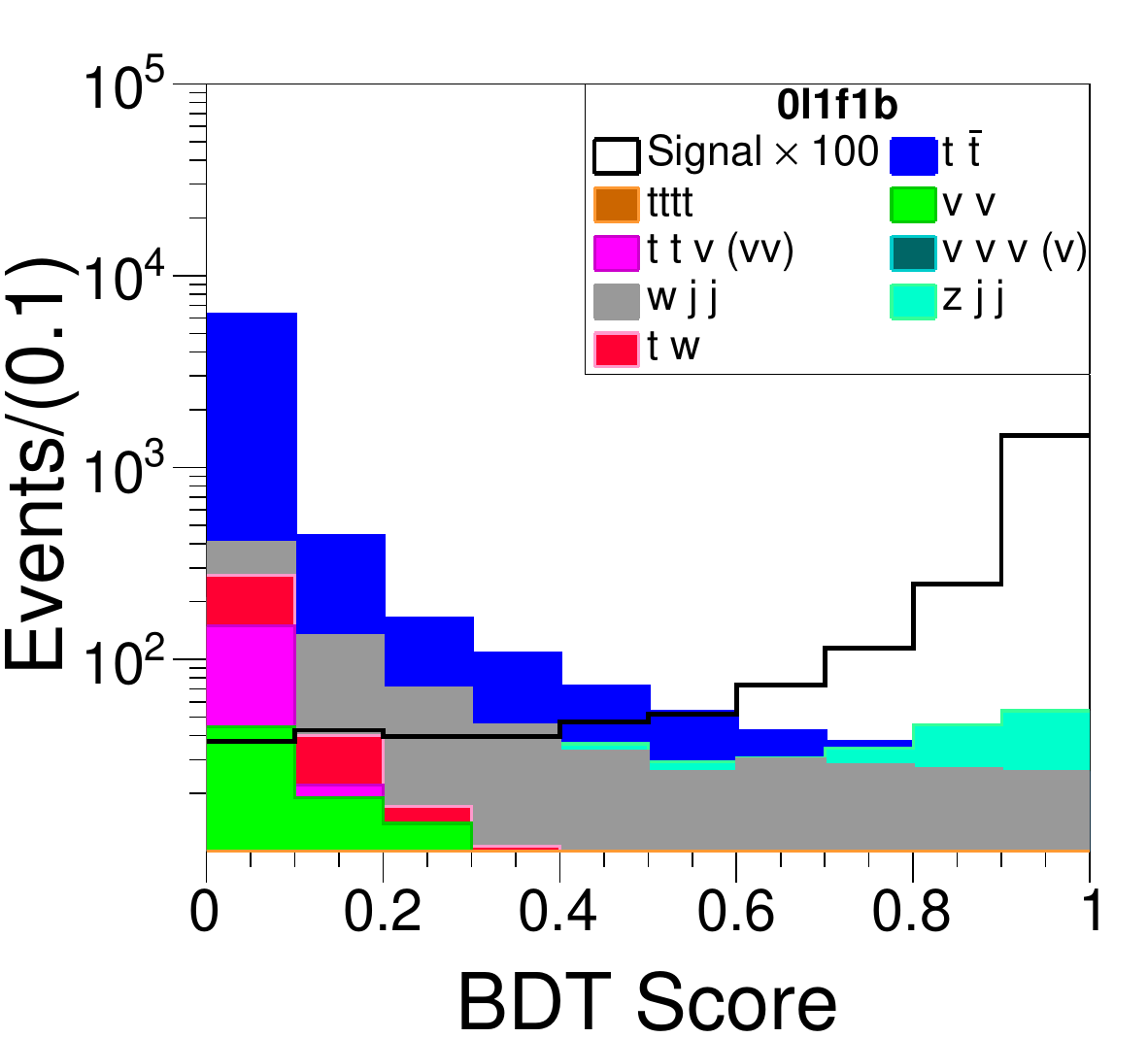}
    \includegraphics[width=0.3\linewidth]{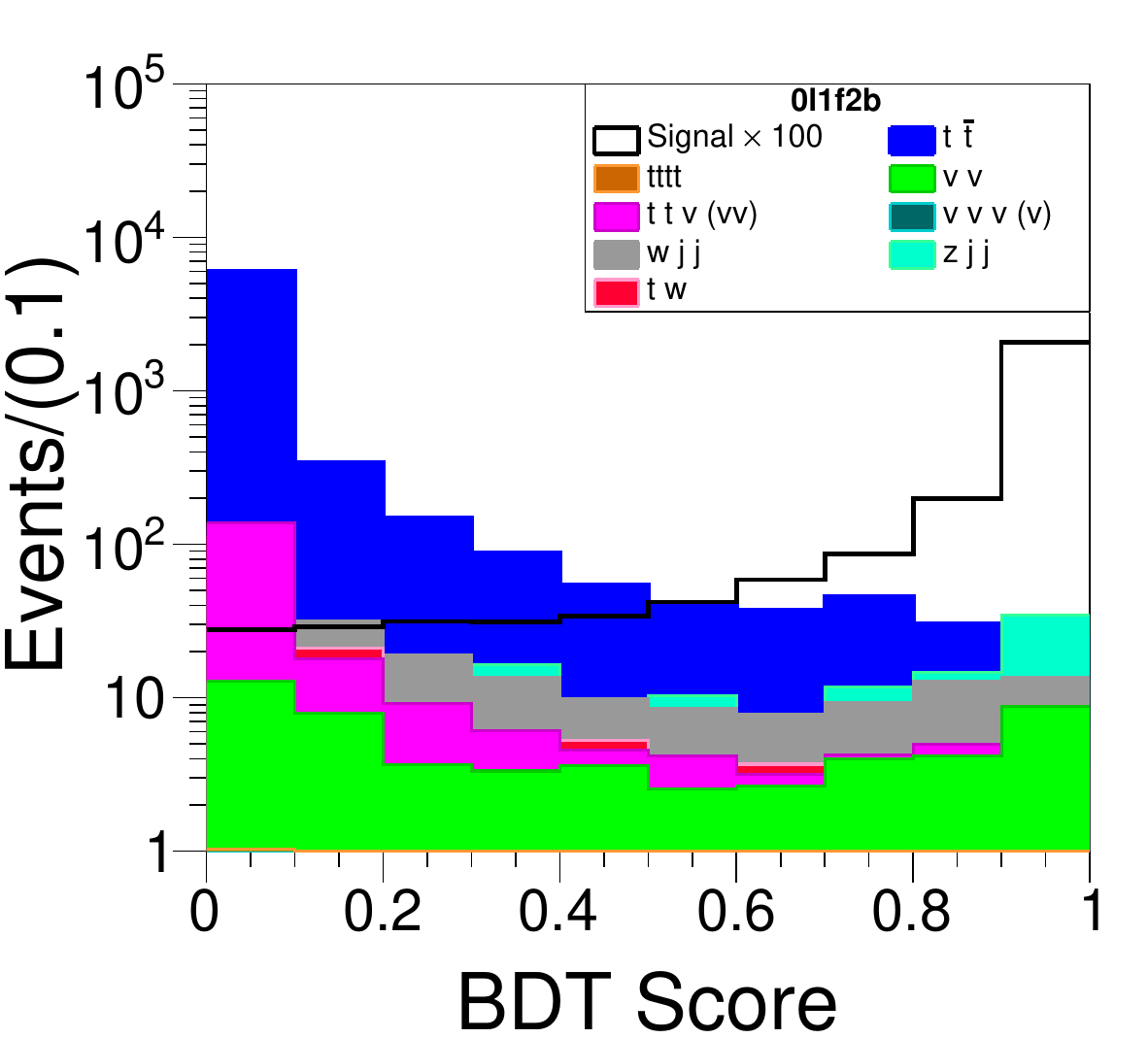}
    \includegraphics[width=0.3\linewidth]{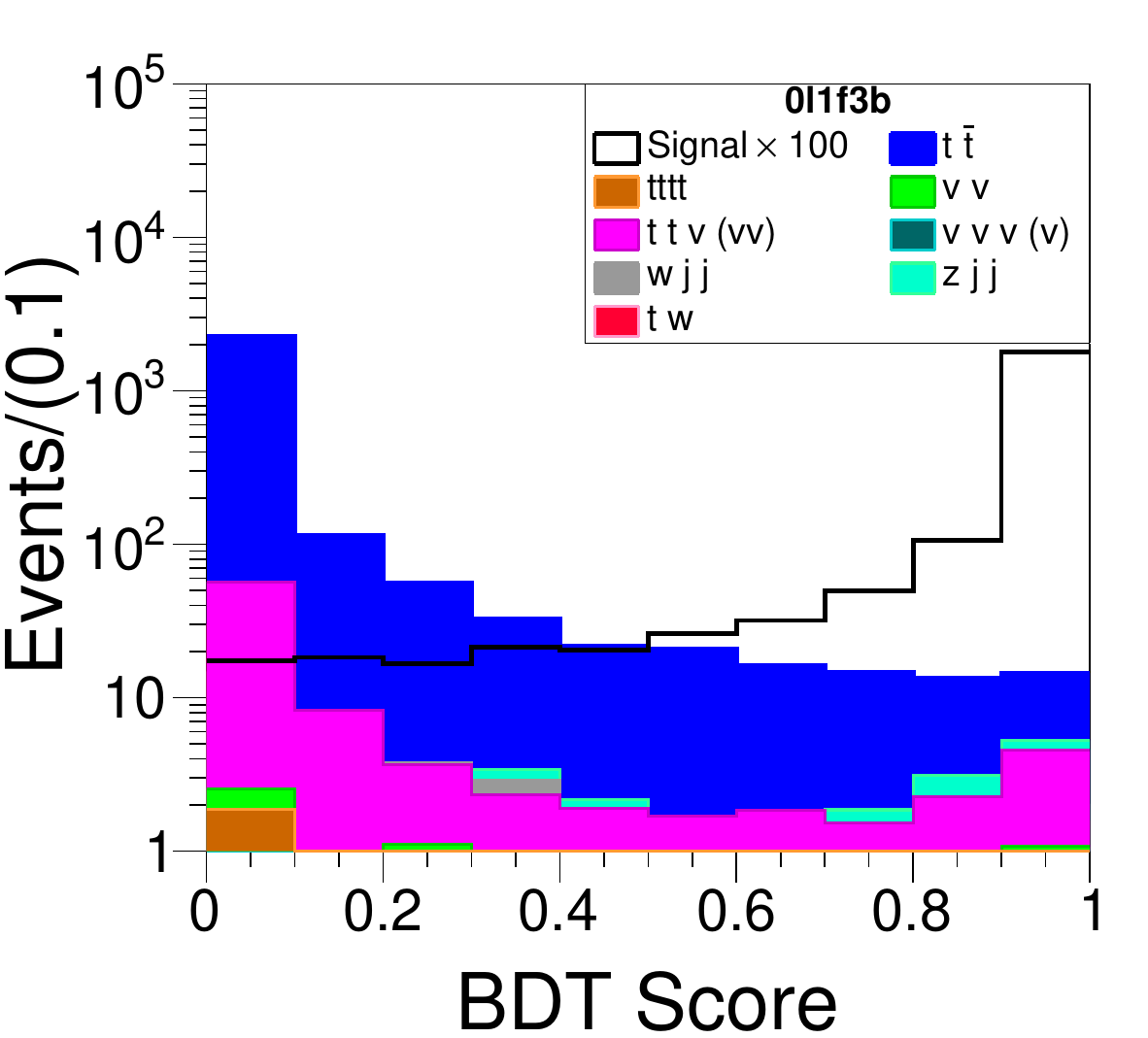}
    \includegraphics[width=0.3\linewidth]{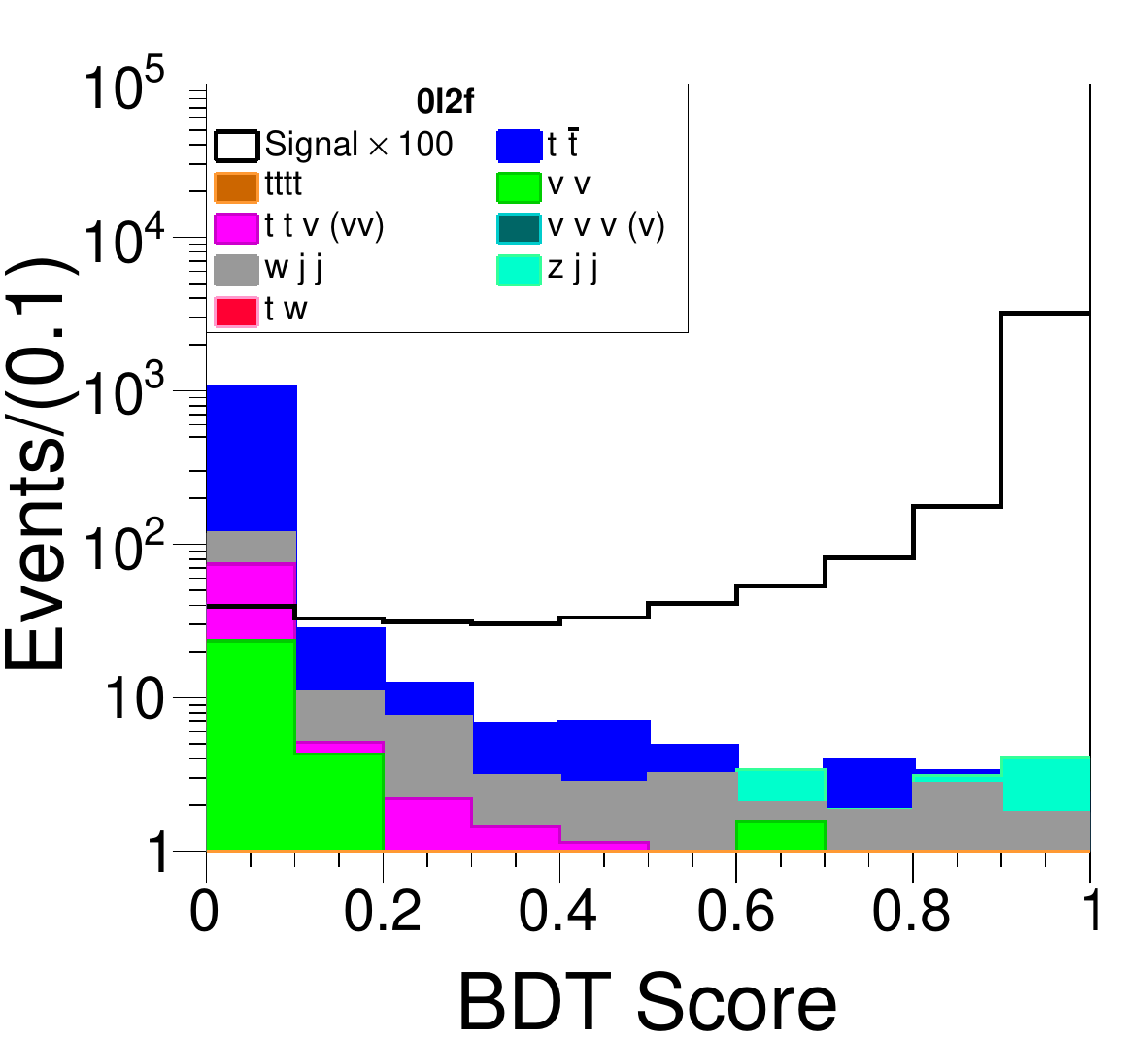}
    \includegraphics[width=0.3\linewidth]{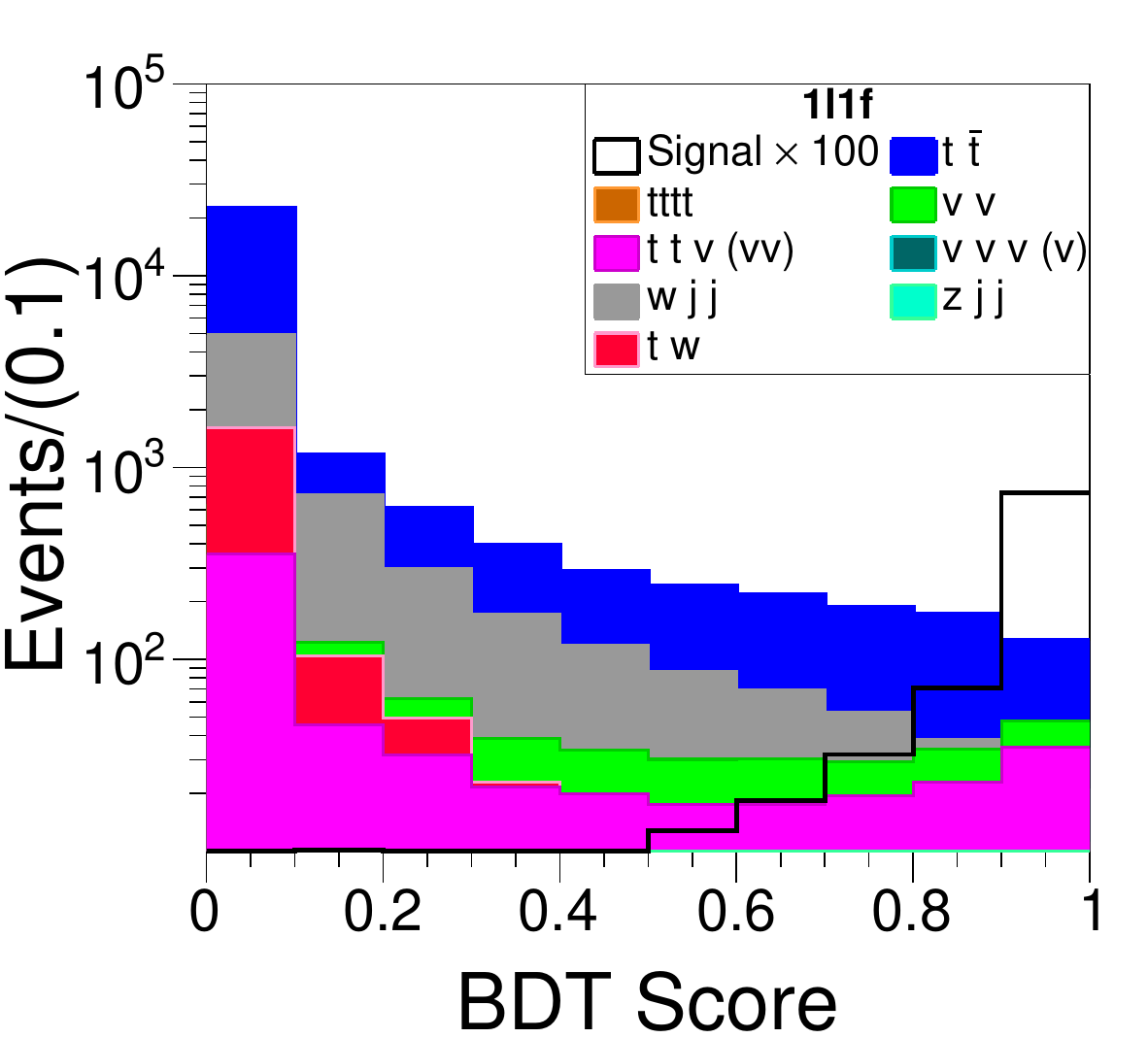}
    \includegraphics[width=0.3\linewidth]{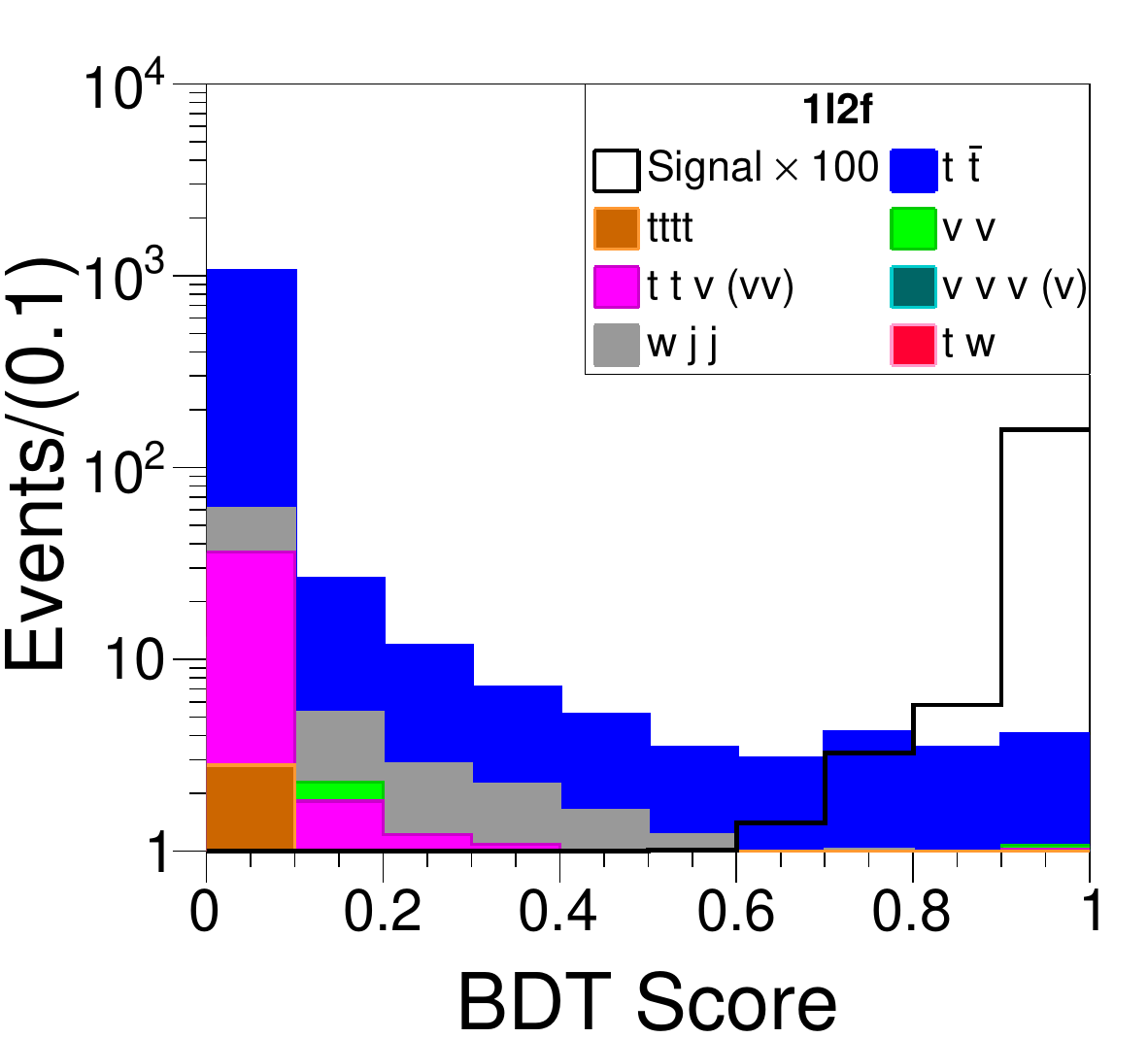}
    \caption{Distribution of BDT score for the seven signal regions for the signal scenario that assumes a 100\% branching ratio of the higgsino into the Higgs boson final state for a signal benchmark point with a 1 TeV higgsino. The y-axis presents the number of signal and background events expected in 200 $fb^{-1}$ integrated luminosity LHC data. For better visibility, we have multiplied the expected number of signal events by a factor of 100. }
    \label{fig:bdtscore}
\end{figure}

In this section, we will focus on the results of our analysis. In Figure \ref{fig:bdtscore}, we present the distribution of the background and signal BDT scores for the seven signal regions for a signal benchmark point with a 1 TeV higgsino. These plots are for the signal scenario that assumes a 100\% branching ratio of the lightest neutralino (in our case, dominantly higgsino) into the $h+\tilde{G}$ final state. We have similar plots for other neutralino masses and the other two values of branching ratios. However, we refrain from presenting them for brevity. \\

%For the 0l1f0b signal region (upper left panel of Figure \ref{fig:bdtscore}), the dominant background contribution comes from the $t\bar{t}$, $Z+j$, and $W+j$ processes with leptonic decay of at least one heavy SM particles providing the necessary missing transverse energy. As we increasingly demand more $b$ jets in the signal region (0l1f1b, 0l1f2b, and 0l1f3b SRs), the contribution from the $Z+j$ and $W+j$ backgrounds progressively reduces while $t\bar{t}$ continues to remain as the dominant background. Similarly, for the 0l2f signal region $t\bar{t}$, $Z+j$, and $W+j$ remain as the dominant background sources as the QCD fat jets can provide the required fat jets as the only requirement imposed is on the invariant mass. For the leptonic signal regions, due to the simultaneous requirement on the number of leptons and missing energy, contribution from $Z+j$ background is reduced significantly, and the  $t\bar{t}$, $t W$, and $W+j$ processes remain as dominant contaminating backgrounds.\\

In the \( 0l1f0b \) signal region (upper left panel of Figure \ref{fig:bdtscore}), the primary background contributions arise from the \( t\bar{t} \) and \( Z/W \) + jets processes. Zero-lepton final states result from these SM processes when the \( W \)-boson decays hadronically and the \( Z \)-boson decays into neutrinos. The SM background events with hadronically decaying \( W \)-bosons are not accompanied by any real source of missing transverse energy. In such events, missing energy arises from the momentum mismeasurement of the final-state particles due to the finite resolution of the LHC detectors. Leptonic decays of the \( W \)-boson can also contribute to the zero-lepton final state when the \( W \)-boson decays into a \( \tau \)-lepton, followed by the hadronic decay of the \( \tau \)-lepton, or when the electron/muon from the \( W \)-boson decay falls outside the coverage of the LHC detectors. As can be seen from the upper left panel of Figure \ref{fig:bdtscore}, the \( Z/W \) + jets process is the dominant SM background for the \( 0l1f0b \) signal region, which is characterized by a \( b \)-jet veto. The background from SM \( t\bar{t} \) production becomes increasingly important as we require a greater number of \( b \)-tagged jets for the signal regions \( 0l1f1b \) (upper middle panel), \( 0l1f2b \) (upper right panel), and \( 0l1f3b \) (middle left panel). For signal regions with at least one electron/muon, namely \( 1l1f \) and \( 1l2f \), the dominant SM backgrounds arise from \( t\bar{t} \) production and the production of \( W \)-bosons in association with multiple jets (see the middle right and bottom panels of Figure \ref{fig:bdtscore}). The simultaneous requirements of electrons/muons and large missing transverse energy effectively suppress contributions from the production of \( Z \)-bosons in association with multiple jets.

Our final analysis treats these BDT scores as the final discriminating observable. After carefully analyzing the signal and background BDT score distribution, we found that a cut on the BDT score at 0.9 works well, reducing the background contribution significantly in each signal region. Then, we split the BDT score between 0.9 and 1 into five bins of size 0.02 each and calculated the expected number of signal and background events in each bin at 200 inverse femtobarn integrated luminosity. Since the seven signal regions are non-overlapping by design, we can statistically combine the expected number of signal and background events for these signal regions to calculate the median expected 95\% CL upper limit on the higgsino pair production cross section. For this part of our analysis, we have used the publicly available tool \textit{pyhf} \cite{Heinrich:2021gyp, Feickert:2022lzh}. \textit{pyhf} is a Python implementation of the \textit{HistFactory} \cite{Cranmer:2012sba} family of statistical models. It provides a flexible\footnote{Unlike \textit{HistFactory}, \textit{pyhf} does not depend on ROOT \cite{Brun:1997pa, Antcheva:2011zz}, which is a complex C++-based system, rather statistical models in \textit{pyhf} are specified using JSON files or dictionaries, which makes them lightweight and easily shareable. It supports multiple computational backends like NumPy, TensorFlow, and PyTorch.} and computationally efficient\footnote{The TensorFlow and PyTorch backend allows the users to use GPUs for intensive computation, thereby improving the computation speed significantly.} framework for hypothesis testing, parameter estimation, and statistical inference. Note that a reliable calculation of upper limits requires a careful estimation of the signal and background uncertainty, which is beyond the scope of our present analysis. We adopt a conservative approach, following the typical LHC searches \cite{CMS:2019lwf, ATLAS:2021eyc}, for which both the theoretical and experimental uncertainties are O(10)\% each, we assume an overall 20\% total signal/background uncertainty.

%For this part of our analysis, we have used the publicly available tool \textit{HistFitter} \cite{Baak:2014wma}. \textit{HistFitter} is a versatile statistical analysis framework used extensively by the ATLAS collaboration for testing various signal hypotheses and setting limits on BSM scenarios. It combines ROOT \cite{Brun:1997pa, Antcheva:2011zz}-based histogram manipulation with statistical techniques from  HistFactory \cite{Cranmer:2012sba}, RooStats \cite{Moneta:2010pm}, and RooFit \cite{verkerke2003roofittoolkitdatamodeling}, allowing for flexible implementation of hypothesis tests, profile likelihood fits, and exclusion limit calculations.

\begin{figure}[!htb]
    \centering
    \includegraphics[width=0.45\linewidth]{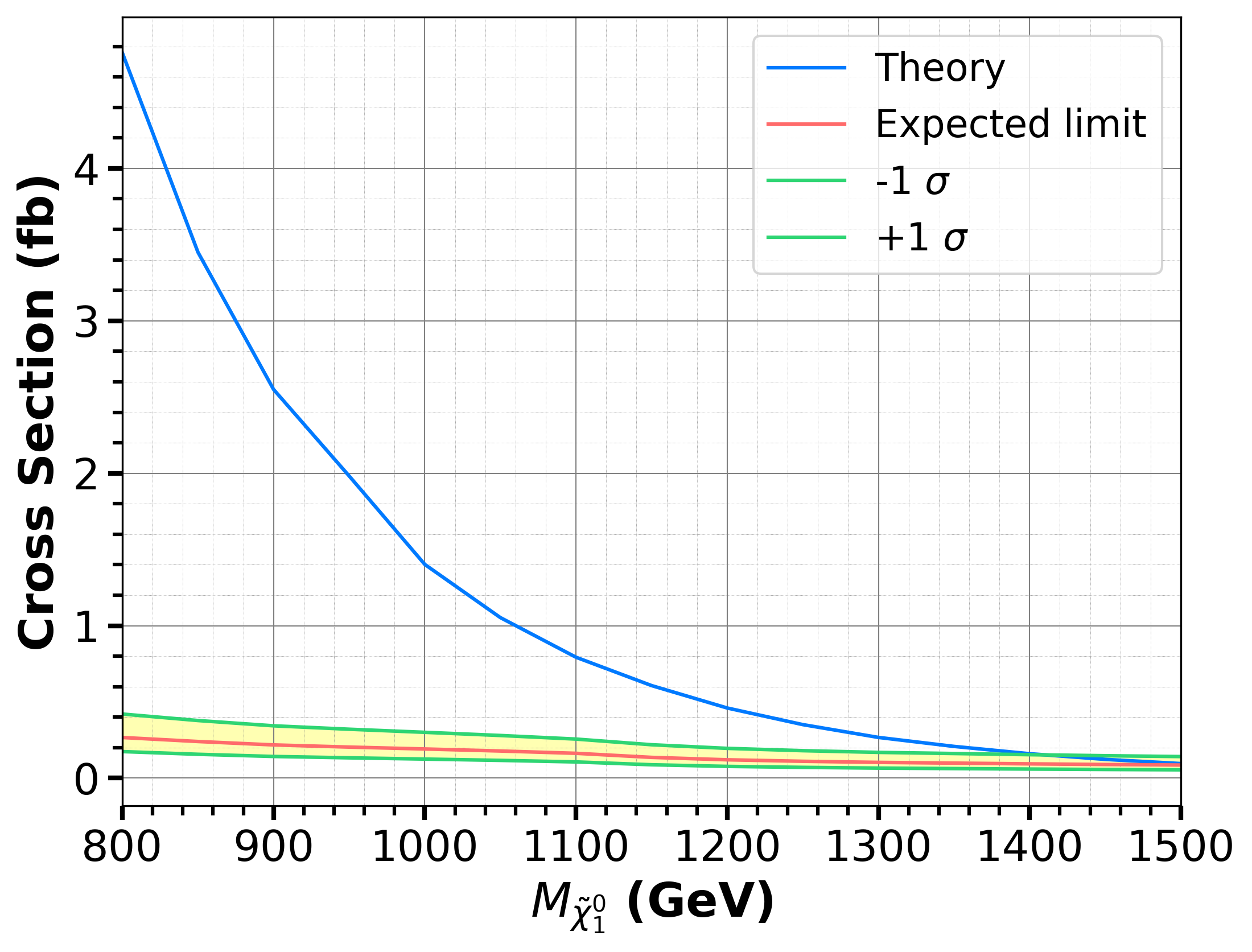}
    \includegraphics[width=0.45\linewidth]{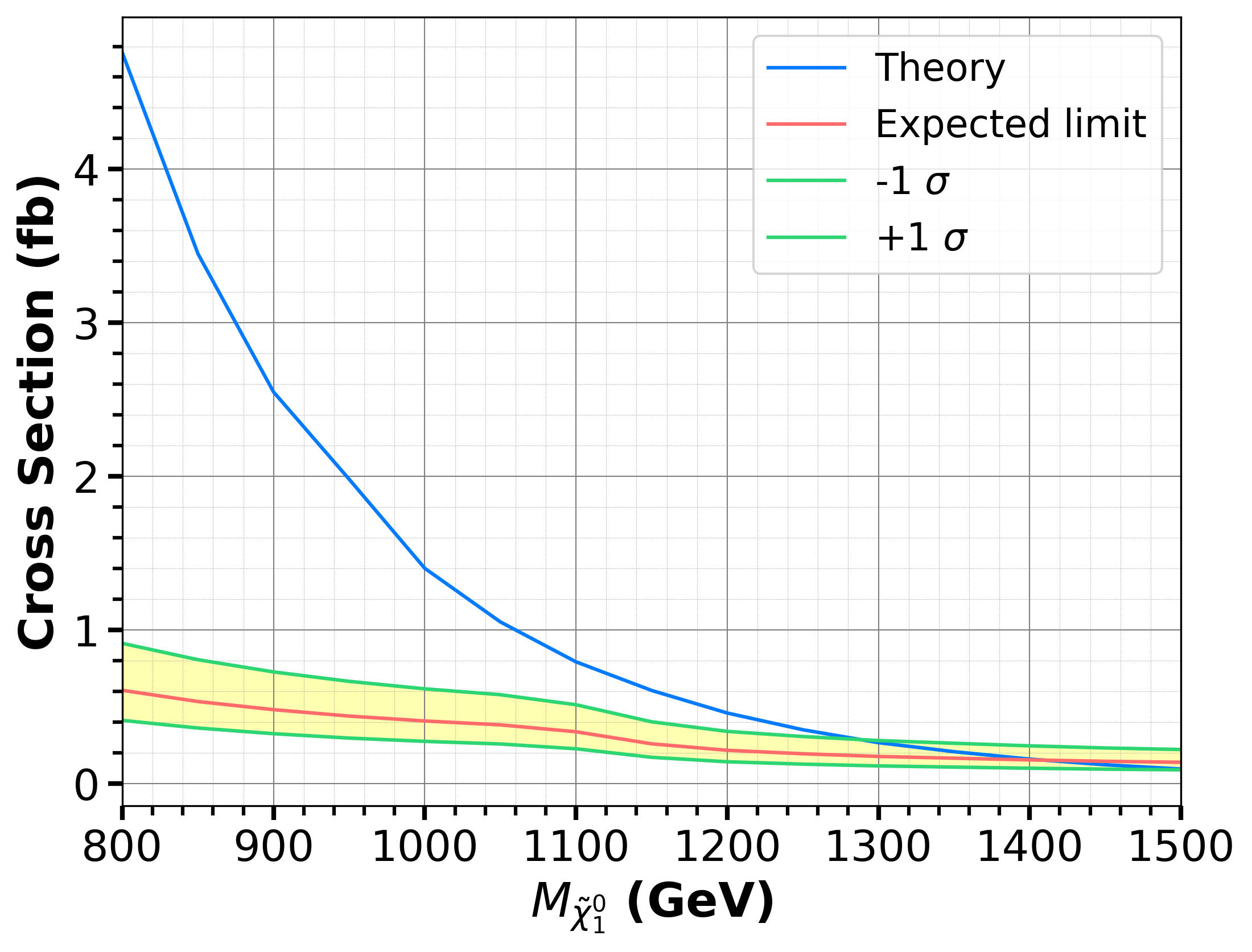}
    \includegraphics[width=0.45\linewidth]{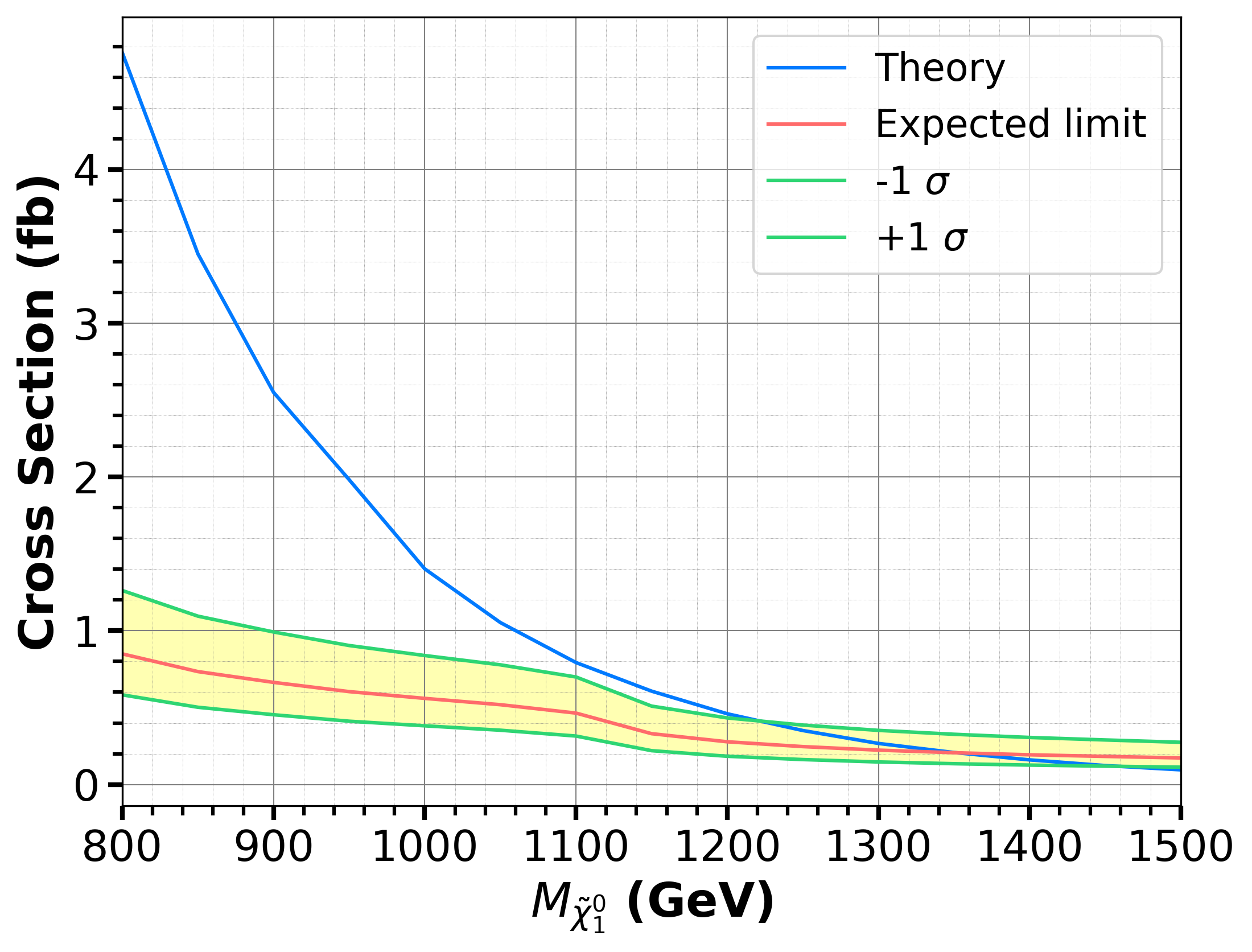}
    \caption{The median expected 95\% CL upper limits on the higgsino pair-production cross-section (solid red line) along with the one sigma uncertainty band (shaded yellow region) for the three signal scenarios considered in our analysis. The plots on the upper left, upper right, and lower panels assume a higgsino decay branching ratio of BR($\tilde{\chi}_1^0 \rightarrow h \tilde{G}) = 100\%$, 50\%, and 25\%, respectively. The NLO+NLL higgsino pair production cross-section is shown by the solid blue line.}
    \label{fig:finalreach}
\end{figure}

Figure~\ref{fig:finalreach} illustrates the median expected 95\% confidence level (CL) upper limit on the Higgsino pair-production cross-section (solid red line), along with the one-sigma uncertainty band (yellow shaded region). The NLO + NLL higgsino pair production cross-section is represented by the solid blue line. The upper left plot corresponds to a branching ratio of BR($\tilde{\chi}_1^0 \rightarrow h \tilde{G}) = 100\%$ and demonstrates that our proposed strategy can exclude Higgsino masses up to \textbf{1470 GeV} at 95\% CL. Similarly, the upper right plot assumes BR($\tilde{\chi}_1^0 \rightarrow h \tilde{G}) = 50\%$, where Higgsino masses up to \textbf{1390 GeV} can be excluded at 95\% CL. The bottom plot considers BR($\tilde{\chi}_1^0 \rightarrow h \tilde{G}) = 25\%$, resulting in the exclusion of Higgsino masses below \textbf{1340 GeV}.

We see a gradual reduction in the reach of our analysis with increasing decay branching ratio of the higgsino into $Z$ boson final state. This behavior can be attributed to two main factors:

\begin{enumerate}
    \item \textbf{Reduced effectiveness of signal regions demanding  $b$-jets}: As BR($\tilde{\chi}_1^0 \rightarrow h \tilde{G}$) decreases, the expected number of $b$-tagged jets in the final state is significantly reduced. Consequently, the signal regions optimized for capturing final states with $b$-jets lose effectiveness, diminishing their contribution to the overall sensitivity.

    \item \textbf{Increased contribution from backgrounds with $W$ fat jets}: With a higher proportion of $Z$ fat jets in the signal events, the contribution from background processes such as $t\bar{t}$ and $W/Z+jets$ becomes more prominent. This is because the multi-class classifier used in our analysis does not distinguish between $W$- and $Z$-initiated fat jets, leading to an increased likelihood of $W$ fat jets from these backgrounds being misidentified as $Z$ fat jets. Consequently, the overlap between signal and background events increases, reducing the overall sensitivity of the analysis.
\end{enumerate}

These two effects, acting in tandem, result in the observed reduction in sensitivity as the branching ratio into $Z$-mediated final states increases.

%The gradual reduction in the reach of our analysis with decreasing decay branching ratio of the higgsino into the Higgs final state is because of the reduction in the expected number of b-quarks in the final state.

\section{Conclusion}
\label{sec:concl}
Light higgsinos with masses ranging from a few hundred GeV to a few TeV are a compelling prediction of natural SUSY models, as they play a crucial role in ensuring the naturalness of the electroweak scale. Extensive searches for these higgsinos within the framework of General Gauge Mediation (GGM) have been conducted by both the ATLAS and CMS collaborations. In a recent study by the CMS collaboration, \cite{CMS:2024gyw} excludes higgsino mass up to 1025 GeV, assuming exclusive decay of the higgsino into the SM-like Higgs boson and gravitino. In this work, we have proposed a novel analysis strategy that leverages the power of graph neural networks (GNNs) through a multiclass classifier to characterize fat jets originating from higgsino decays. As we only expect SM $Z$ or Higgs boson fat jets in our signal events, this procedure gives us an extra handle in suppressing the dominant SM backgrounds lacking these fat jets. By carefully designing signal regions and kinematic variables tailored to the expected nature of the signal events in these signal regions, we found that a boosted decision tree classifier can probe higgsino masses up to 1470 GeV, 1390 GeV, and 1340 GeV at the 14 TeV LHC with 200 $fb^{-1}$ integrated luminosity proton-proton collision data, assuming decay branching ratios of 100\%, 50\%, and 25\% into the Higgs-mediated channel, respectively.

Contrary to conventional search strategies employing neural network (NN) based fat jet classifiers, our approach achieves an improved signal yield. Traditional strategies typically attempt to identify the parent particle of fat jets by imposing a threshold on the neural network (NN) score. However, these thresholds are associated with specific identification/tagging efficiencies, resulting in a loss of signal events during the identification process. In contrast, our novel strategy avoids imposing strict thresholds on the GNN scores of fat jets. Instead, the GNN scores are treated as variables that characterize the parent particles of the fat jets. This approach ensures that all signal events remain available for the final analysis, where a secondary boosted decision tree classifier utilizes these GNN scores alongside additional high-level variables to distinguish signal from background events. Similar ideas, where jet constituent information is combined with additional event-level variables (using a transformer encoder with a cross-attention layer) without explicit fat jet identification, have been explored in Ref. \cite{Hammad:2023sbd}. However, unlike the fixed number of fat jets required by the transformer architecture in Ref. \cite{Hammad:2023sbd}, our approach allows the number of fat jets used in the final analysis to vary dynamically from event to event, making it more flexible and adaptable to diverse scenarios.

As a proof of concept, we applied this strategy to probe higgsinos within the GGM framework and demonstrated that it offers better sensitivity than prior LHC searches. Importantly, this methodology is highly general and can be applied to a wide range of collider searches focusing on boosted final states. Potential applications in GGM-type SUSY searches include scenarios where the neutralino is a bino-higgsino admixture, leading to $h/Z$ bosons in the final state, GGM scenarios with wino-co NLSPs producing $W$ and $Z$ bosons, and those with chargino NLSPs yielding $W$ bosons. While a detailed exploration of these scenarios lies beyond the scope of this study, we aim to address them in future work.

\section{Acknowledgement}
KG acknowledges ANRF India for providing Core Research Grant no. CRG/2023/008570.

\appendix

\section{Multiclass classifier}
\label{sec:gnn_model}
Our analysis uses a GNN-based multiclass classifier to identify the boosted parent particles resulting in fat jets. These fat jets can either originate from the decay of heavy SM particles like the top quark, Higgs boson, and W/Z boson or from the hadronization of QCD-initiated light quarks or gluons (hereafter QCD jets). Due to the similarity in mass and hadronic decay topology, we consider fat jets originating from W and Z bosons to be a single class called V jets. Accordingly, we train our multiclass classifier to predict the probability with which a fat jet belongs to these four classes.\\

For the GNN model, we used the publicly available architecture LorentzNet \cite{Gong:2022lye}. Note that the original LorentzNet architecture was designed for a binary classification problem. To use it for our analysis, we have changed the number of nodes in the output layer to suit our multi-class classification task. Recognizing that the original architecture was optimized for binary classification, we conducted an extensive hyperparameter scan to identify configurations best suited for multi-class classification. Through this process, we determined that incorporating four Lorentz group equivariant blocks (LGEBs) yields optimal performance for our problem.  Beyond these adjustments, we found that the core design of LorentzNet—such as the design of the LGEBs, input block, and decoding block (see Ref.\cite{Gong:2022lye} for details)—aligned well with the requirements of our analysis, and no further substantial changes were necessary.\\

To generate fat jets for training and testing the classifier, we followed a strategy similar to that outlined in Ref. \cite{Gong:2022lye}. Top, W/Z, and QCD samples are generated from the SM processes $p p \rightarrow t \bar{t}$, $p p \rightarrow W^+ W^-/ ZZ$, $p p \rightarrow j j$ (where j includes u,d,c,s, their anti-particles, and gluon), respectively. To generate Higgs fat jets, we used Higgs pair production using the default \textbf{\textit{heft}} model of madgraph. The rest of the data generation procedure is similar to that of Ref \cite{Gong:2022lye}, with a few minor changes. First, instead of using the constituent masses as the node embedding scalar, our analysis uses their charge. Second, unlike Ref. \cite{Gong:2022lye}, we generate fat jets over a wide transverse momentum range from 300 to 1500 GeV. To ensure a uniform population of fat jets over this wide range, we break the transverse momentum into 50 GeV bins and generate 50k fat jets from each of the four categories for training, 25k for testing, and 25k for validating the classifier. For a detailed discussion on the performance of the classifier, we urge the interested readers to consult Ref. \cite{leptoquark}.

\section{Kinamatic Variables}
\label{sec:kinametic_variables}
This section provides a detailed discussion of the kinematic variables constructed to train the BDT classifiers for each of the seven signal regions. The variables used in the analysis are as follows:\\
$p_{T,i}$, $\eta_i$, $S_{h,i}$, $S_{t,i}$, $S_{V,i}$, $\Delta \phi (i,j)$, $\not\!p_T$, $M_{eff}$,$H_T$, $\not\!p_T/\sqrt{H_T}$, $ (p_{T,j_1}+p_{T,j_2})/H_T$, $M(j_i,j_j)$, $\Delta \phi (\not\!p_T,i)$, $N_{b \in j_i}$, $N_{b \in fj_i}$, $N_{b \in l_i}$, $\sum _{j\notin b} (p_T(j))$, $\sum _{j\in b} (p_T(j))$

To clarify our notation, the following conventions are adopted:

\begin{enumerate}
    \item In $p_{T,i}$, $\eta_i$, and $\Delta \phi (i,j)$,  $i$ and $j$ stand for all the signal region particles, including the leptons, fat jets, and light jets. Note that we denote by the symbol $j$ all the small radius jets, including the b-tagged jets, and the symbol $b$ is exclusively used for the $b$-tagged jets. 
    \item $M(j_i,j_j)$ stands for the invariant mass of all possible light jet pairs.
    \item We have defined $H_T$ as the scalar sum of the transverse momentum of all visible final state particles. $M_{eff}$ is defined as the sum of $H_T$ and $\not\!p_T$. 
    \item $S_{h,i}$, $S_{t,i}$, and $S_{V,i}$ denote the h-score, t-score, and v-score of the fat jets, respectively. The index $i$ runs over all the fat jets in the signal region.
    \item $N_{b \in j_i}$, $N_{b \in fj_i}$, and $N_{b \in l_i}$ stand for the number of b jets within a cone of radius 1.2 of a light jet, fat jet, and lepton, respectively.
    \item $\sum (p_T(j\notin b))$ represents the scalar sum of the transverse momentum of light jets not tagged as b-jets.
\end{enumerate}

\section{BDT Hyperparameters}
\label{sec:hype_params}
In Table \ref{tab:xgboost_settings}, we present the hyperparameter setting of the XGBOOST classifiers.

\begin{table}[h!]
\centering

\begin{tabular}{|l|l|}
\hline
\textbf{Parameter}          & \textbf{Value}                \\ \hline
Objective                   & \texttt{binary:logistic}      \\ \hline
Maximum depth (\texttt{max\_depth}) & 4                          \\ \hline
Learning rate (\texttt{eta})       & 0.01                       \\ \hline
Evaluation metric (\texttt{eval\_metric}) & \texttt{logloss}           \\ \hline
Seed              & 42                              \\ \hline
Number of boosting rounds (\texttt{num\_boost\_round}) & 700 \\ \hline
\end{tabular}
\caption{Hyperparameter settings for training the XGBoost classifier.}
\label{tab:xgboost_settings}
\end{table}

\newpage
%=========================
\bibliographystyle{JHEP}
\bibliography{v0}
%==========================

\end{document}